\documentstyle[eqsecnum,pre,preprint,aps]{revtex}
\begin{document}

\draft
\title{Correlations between eigenvalues of large random matrices
         with independent entries}
\author{J. D'Anna$^1$ and A. Zee$^{1,2}$}
\address{$^1$ Department of Physics,University of California,
        Santa Barbara, California  93106-9530 \\
        $^2$ Institute for Theoretical Physics, University of California,
        Santa Barbara, California  93106-4030}
\date{\today}
\maketitle

\begin{abstract}
We derive the connected correlation functions for eigenvalues of large
Hermitian random matrices with independently distributed elements using
both a diagrammatic and a renormalization group (RG) inspired approach.
With the diagrammatic method we obtain a general form for the one,
two and three-point connected Green function for this class
 of ensembles when matrix elements are identically distributed,
 and then discuss the derivation of higher order functions by the same
approach.
Using the RG approach we re-derive the one and two-point Green functions
 and show they are unchanged by choosing certain ensembles with non-identically
distributed elements. Throughout, we compare the Green functions we obtain to
those from
the class of ensembles with unitary invariant distributions and discuss
universality in
both ensemble classes.\\
\end{abstract}

\section{Introduction}
\label{sec:intro}

Of recent interest in random matrix theory has been the discovery and
characterization of new universal
behavior in the densities of, and correlations between the eigenvalues of
various ensembles which may be relevant to the study of mesoscopic systems,
disordered metals, and random surfaces
\cite{bz1,bz2,bz3,bz4,bzj,bee,for,eynard}.

In this paper
we present an investigation for the existence of universal behavior in
 the correlation functions of a particular class of random matrix ensembles,
those comprised of real symmetric or Hermitian  matrices with individually
distributed random elements.
 A notable example of
an ensemble in this class is the first random matrix ensemble
considered by Wigner \cite{wig}, {\sl i.e.} real symmetric matrices
whose elements assume the values
$\pm\sigma/\sqrt{N}$ with equal probability.  Here we will refer to
this class of theories as the ``Wigner class".

It has been known for quite some time that in Wigner class ensembles the
eigenvalue
density is universal.
More specifically, the averaged density of eigenvalues for an ensemble of
matrices with
individually identically distributed elements is the same
for any chosen distribution, up to a scaling of the width of the spectrum.
We define the averaged density of eigenvalues as
\begin{equation}
\rho(\lambda) =
\left\langle {1\over N}\text{Tr}\delta(\lambda-\varphi) \right\rangle
\end{equation}
 where $\varphi$ is an $N\times N$ matrix and $\langle . \rangle$
denotes averaging over an ensemble of such matrices.  In general, the ensemble
average
for an operator, ${\cal O}$, is obtained via the integral
\begin{equation}
\langle{\cal O}(\varphi)\rangle \equiv \int d\varphi P(\varphi) {\cal
O}(\varphi)
\end{equation}
where $P$ is a density function, to be defined in more detail later,
which takes as its argument the matrix $\varphi$.
For any  Wigner Class ensemble one obtains the well known result
\begin{equation}
\rho(\lambda) = {2 \over \pi a^2} \sqrt{a^2 - \lambda^2}, \label{ssl}
\end{equation}
which is known as Wigner's ``semi-circle'' law.

In this paper we will  compare our results  for ``Wigner class" ensembles to
those of
another class, ensembles specified by a matrix density invariant under a
symmetry group
transformation.  These ``invariant class" (or ``trace class'' in the
terminology of Refs.  \cite{bz1,bz2,bz3,bz4,bzj}) ensembles
have been shown to have
non-universal eigenvalue densities.  In particular, one finds the density of
eigenvalues for an
invariant class ensemble may be written
\begin{equation}
\rho(\lambda) = {2 \over \pi a^2}p(\lambda) \sqrt{a^2 - \lambda^2},
\label{mssl}
\end{equation}
with $p(\lambda)$ a polynomial which depends on the ensemble distribution in a
complicated way \cite{bpiz}.  However, the two-point and higher, connected
correlation functions for this class have been found to be universal
in form  in the limit that
$N \rightarrow \infty$, where $N$ denotes the dimension of the matrices
\cite{bz1,bz2,bz3,bz4,bee}.  Their only dependence on the averaging ensemble is
through the endpoint of the eigenvalue spectrum.

In a previous work \cite{bz3}, Br\'ezin and Zee suggested that the two-point
and higher correlation functions for the Wigner class (in the
$N\rightarrow\infty$ limit)
were universal as well, having the same form as those for the invariant class.
Here we will show that this suggestion was incorrect, and that the two-point
correlation
function is somewhat less universal than its invariant class analog, as it
depends on two parameters of the distribution (to be defined below) rather than
just
one.  In general the averaged $k$-point connected correlation for a Wigner
class
ensemble will explicitly depend on the ensemble's
first $2k$ moments, and will be a quantity of order $N^{-3k/2+1}$ to
$N^{-2k-2}$
depending on which moments are present.

Amusingly, the occurrence of universality is
reversed between these two classes.  While the eigenvalue density for the
Wigner
class is universal, the higher order connected correlation functions
increasingly depend on the particulars of the ensemble distribution chosen.
In contrast, in the invariant class the eigenvalue density is not universal,
but the
higher order connected correlation functions are.

For the two-point connected correlation function in the Wigner class
(which we calculate for the Hermitian matrix case in the following sections)
we find an expression dependent on only
the second and fourth moments of the averaging ensemble.  Specifically, we
obtain
\begin{equation}
\rho_c(\mu,\nu) =
{-1\over 2\pi^2 N^2}
\left(
{(4\sigma^2-\mu\nu) \over\beta (\mu-\nu)^2
        \sqrt{(4\sigma^2-\mu^2)(4\sigma^2-\nu^2)}}
-{\tau^4\over4\sigma^8}
        {(4\sigma^2-2\mu^2)(4\sigma^2-2\nu^2)
        \over\sqrt{(4\sigma^2-\mu^2)(4\sigma^2-\nu^2)}}
\right),
\label{2ptfunc}
\end{equation}
where $\beta = 1$, or $2$ depending on if one considers real orthogonal or
Hermitian
 matrices, $\sigma^2$ is the second moment of the ensemble distribution,
$\tau^4$
 the fourth moment, and $\pm2\sigma$ are the endpoints of the eigenvalue
spectrum,
 all to be more explicitly defined later.
Within the Wigner class
one may consider, for example, such arbitrary and disparate ensemble
distributions
 as $P_N(\varphi_{ij})\propto \Theta(c^2/N - |\varphi_{ij}|^2)$,
$P_N(\varphi_{ij})\propto \exp(-N\sum^p_{k=1}g_k|\varphi_{ij}|^{2k})$, or
$P_N(\varphi_{ij})\propto
1/2[A\delta(\varphi_{ij}-c/\sqrt{N})+B\delta(\varphi_{ij}+c/\sqrt{N})]$
always obtaining Eq.(\ref{2ptfunc}).

If we set $\tau^4=0$  in Eq.(\ref{2ptfunc}) we
obtain the result for a Gaussian ensemble distribution.  This is identical to
the universal form
of the two-point connected correlation that was obtained by Br\'ezin and Zee
for any distribution in the invariant class.  The Gaussian case, of course,
belongs to both the Wigner and invariant classes.  However, in general the
two-point correlation for the Wigner and the invariant class differ by an
additional term, namely the second term in Eq.(\ref{2ptfunc}).

In Fig.\,\ref{theory1} and Fig.\,\ref{theory2} we have plotted the functions
$(\mu-\nu)^2\rho_c(\mu,\nu)|_{\beta=1}$ and
$(\mu-\nu)^2\rho_c(\mu,\nu)|_{\beta=1,\tau^4\rightarrow 0}$
with the arbitrary choice  $\sigma^2 = 1$, $\tau^4 = -2$, and $N=100$
for two specific values
of $\mu$. We see that the second term is capable of changing the shape of the
curves substantially.  Thus, two Wigner systems with identical eigenvalue
densities may have very different two-point connected eigenvalue correlations.

In a numerical work Kobayakawa, {\sl et. al.}\cite{khkz}
found poor agreement between numerical calculations
of two-point correlations for specific Wigner ensembles and the universal
form for the invariant class.
It appears that, with the inclusion of the second term, Eq.(\ref{2ptfunc}) is
 in good agreement with the numerical results \cite[see Figures \ref{khkz1} and
\ref{khkz2}
which correspond to (19) and (20) in
Ref.]{khkz}.

Recently, we have learned that Pastur, Khorunzhy, and Khoruzhenko
\cite{past} have also calculated correlation functions for the general
Wigner class, but using a  completely different technique.  We feel it
 worthwhile to present an independent study, since the techniques
we use may be of interest, and the form of the resulting equations we obtain
elucidating.
Our result (\ref{2ptfunc}) differs from theirs (Eq.(14) in Ref.\cite{past}) in
the sign of the second term.

This article is organized as follows.  In Sec.\,\ref{sec:diag} we employ a
diagrammatic
approach to calculate the averaged one, two and three-point Green functions,
or resolvents, for general Wigner class ensembles in the  large $N$  limit.  We
then
outline how all higher point Green functions may similarly be calculated.
In Sec.\,\ref{sec:rg} we demonstrate a completely different approach to the
same problem, namely a renormalization group (RG) inspired procedure for
calculating
the same Wigner class Green functions.
Naturally, the methods of both sections return identical results.

Since from the diagrammatic standpoint this problem takes
the form of a large $N$, zero dimensional field theory calculation, it is
straightforward to determine the relevant diagrams and sum them.
The RG method, on the other hand, takes more work and
proves less transparent. However, the RG method makes more  apparent the
freedom
one has in choosing  model ensemble densities without changing the resulting
correlation
functions.

\section{correlations via diagrams}
\label{sec:diag}

The use of diagrammatic methods in the study of random matrix theories
is not new.  Such techniques have been successfully applied in both invariant
class
and Wigner class calculations\cite{bz3,verb}  with applications to
the latter being limited to a few specific but important ensembles,
namely the ones with Gaussian distributed elements.

Here we will
show one how may calculate via diagrams the hierarchy of correlation
functions for the Wigner class in general.  The known results for Gaussian
ensembles will be obtained as a special case.  We will begin by explicitly
calculating
the one and two-point functions before considering the
general $n$-point case.

Let $\varphi$ be an element of an ensemble of dimension $N$ matrices with
individually identically distributed elements. We limit our consideration
to Hermitian matrices, however the method may be extended
to other sets such as symplectic or real symmetric.  Also, we will initially
take the case of an even distribution, and save treating the inclusion
of odd moments until Sec.\,\ref{sec:3pthigher}.

We choose to define the moments of the matrix element distributions so that
the eigenvalues of any matrix will be of order $N^{-1}$ with mean zero
and range of order $N^0$.  If we write $\varphi = X + iY$, with
$X^i_j= X^i_j$, and $Y^i_j = -Y^j_i$ this is accomplished by defining:
$\langle X^i_j\rangle=\langle Y^i_j\rangle=0$,
$\langle( X^i_j)^2\rangle = (1+\delta^i_j)C_2/2N$,
$\langle( Y^i_j)^2\rangle = (1-\delta^i_j)C_2/2N$,
{\it etc.} so that
$\langle\varphi^i_j\rangle =0$,
$\langle\varphi^i_j\varphi^{i*}_j\rangle =  C_2/N$,
$\langle(\varphi^i_j\varphi^{i*}_j)^2\rangle - 2(C_2/N)^2 = C_4/N^2$,
{\it etc.},
where each $C_n$ is some constant of order $N^0$ and $\langle . \rangle$
denotes averaging over the ensemble. Note that in the case of a Gaussian
ensemble all
 moments but $C_2$ equal zero.

In general the $k$th moment is $C_k/N^{k/2}$. The inverse powers of $N$ appear
as a result of our requirement that the eigenvalue spectrum be finite in width
as
$N$ becomes large. Alternatively we might have defined the ensemble density as
having
moments $C_2,C_4,\ldots C_k$, and the matrices of the ensemble as having
elements
$\varphi^i_j = X^i_j/\sqrt{N} + i Y^i_j/\sqrt{N}$, with the same effect. We
have
opted, however, to define the distribution as above in order to facilitate
 our later counting of the powers of $1/N$ associated with each diagram.

We define the matrix
\begin{equation}
\hat{G}_N(z)^i_j \equiv \left( {1\over z - \varphi}\right)^i_j, \label{ghat}
\end{equation}
and the Green function, or resolvent, as
\begin{equation}
G_N(z)   \equiv
\left\langle {1\over N}\text{Tr}\,\hat{G}_N(z) \right\rangle. \label{gn}
\end{equation}
We are interested in $G(z) \equiv \lim_{N\rightarrow\infty} G_N(z)$
from which we may extract the averaged eigenvalue density via
$\rho(\mu)\left.-{1\over\pi}%
\text{Im}\,G(\mu+i\epsilon)\right|_{\epsilon\rightarrow 0}$.

Similarly we define the two-point Green function as
\begin{equation}
G_N(w,z)  \equiv  \left\langle {1\over N}\text{Tr}\,\hat{G}_N(z) {1\over N}
\text{Tr}\,\hat{G}_N(w)\right\rangle.   \label{g2n}
\end{equation}
In the large $N$ limit $G_N(w,z) \rightarrow G(w) G(z)$; thus we define the
{\it
connected} function as
\begin{equation}
 G_{cN}(w,z) \equiv G_N(w,z) - G_N(w) G_N(z).
\label{g2c}
\end{equation}
We see that this will be a quantity of order $N^{-2}$.
The connected two-point correlation may then obtained via
\begin{equation}
\rho_c(\mu,\nu) = -{1\over 4\pi^2}[G_c(+,+)+G_c(-,-)-G_c(+,-)-G_c(-,+)],
\label{2ptdif}
\end{equation}
where $G_c(\pm,\pm) \equiv \left. G_c(\mu\pm i\epsilon,\nu\pm
i\delta)\right|_{\epsilon,\delta\rightarrow 0}$.
Three-point and higher Green functions may be defined, and their corresponding
correlation functions obtained in a similar fashion.

\subsection{One point Green function}

To calculate $G(z)$ we consider the untraced average of $\hat{G}(z)$ written
 as a power series expansion in $1/z$, and then evaluate the ensemble
average of each term in the series.  Thus, we start with
\begin{equation}
G_N(z)^i_j = \sum_{n=0}^\infty
        {1\over z}\left\langle\left[\left(\varphi {1\over
z}\right)^n\right]^i_j\right\rangle
\label{gseries}
\end{equation}
The average of any given term in the series may be obtained by
summing over all the possible ways of taking the averages of groups of
$\varphi$'s---
pairs, triplets, {\it etc.}
For example, the $n=4$ term of Eq.(\ref{gseries}) is given by,
\begin{equation}
{1\over z}\left\langle\left(\varphi{1\over z}\right)^4\right\rangle
={1\over z}\hat{\varphi}{1\over z}\hat{\varphi}{1\over z}\tilde{\varphi}{1\over
z}\tilde{\varphi}{1\over z}
+{1\over z}\hat{\varphi}{1\over z}\tilde{\varphi}{1\over z}\hat{\varphi}{1\over
z}\tilde{\varphi}{1\over z}
        +{1\over z}\hat{\varphi}{1\over z}\tilde{\varphi}{1\over
z}\tilde{\varphi}{1\over z}\hat{\varphi}{1\over z}
+{1\over z}\hat{\varphi}{1\over z}\hat{\varphi}{1\over z}\hat{\varphi}{1\over
z}\hat{\varphi}{1\over z},
\label{wigcontractions}
\end{equation}
where the constituents of a group average are indicated by having either a
tilde or hat above them.
Given our choice that $\langle\varphi^i_j\rangle=0$,
single and triple $\varphi$ averages do not contribute to this
particular term.
It should be emphasized that this procedure entails no approximation

        In order to evaluate (\ref{gseries}) by this procedure
we consider it as a diagram expansion and use the double line
formalism of 't Hooft \cite{thoo},
where the lines in the diagrams correspond to the indices of the matrices
 being averaged. Figure \ref{4ds} shows the diagrams associated with the
terms in Eq. (\ref{wigcontractions}).
Each single line corresponds to a $\delta^i_j/z$
and each double line connects a $\varphi^i_j$ to one or more others in an
average. To
borrow from the particle physics literature we might call these ``quark
propagators"
and ``gluon propagators" respectively.
Then two $\varphi$'s averaged together correspond to an
emitted and absorbed gluon.  The higher moments of a particular distribution
correspond to gluon ``interaction vertices" such as the four
gluon vertex depicted in Fig. \ref{4ds}(d) \cite{note}.

        The great advantage of the double line formalism is that it allows one
to count the powers of $N$ associated with each diagram with ease.
Each continuous closed line represents a sum over the
values of an index, and contributes a factor of $N$.
Thus, non-planar diagrams, such as Fig. \ref{4ds}(b),
are suppressed by powers of $1/N$ relative to equivalent planar
ones since a non-planar diagram always has fewer closed index sums
than a planar diagram having the same ``interactions".

        The nature of the gluon--gluon interaction vertices affords another
simplification
in the large $N$ limit. Each $k$th order gluon interaction
vertex, such as the fourth order one in Fig. \ref{4ds}(c),
contributes a factor of $C_k/N^{k/2}$ (in this way of counting the powers of
$1/N$
associated with the gluon propagators are included).
However, any such vertex has only two indices which may be summed over.
Figure \ref{vertices}(a) depicts a four-gluon vertex with lines labeled
by their index.
If such a vertex appears in a diagram,
that diagram will have a contribution of order $N^{-(k/2-1)}$ , or less,
to the total expansion.

This is the crucial, and simple, reason
 why the one-point Green function for this class of ensemble is
universal. In the large $N$ limit the expansion of  Eq.\,(\ref{gseries})
only depends on diagrams containing averages of pairs of matrices.
In particle physics language we would say that the interactions between
the gluons becomes weaker and weaker as $N$ approaches infinity.
Thus for large $N$, regardless of
the particular features of a distribution, the one-point Green function and
averaged
eigenvalue density depend on only one number: the ensemble density's
second moment.

        This is in direct contrast with the result for
invariant class ensembles.  Invariant
class one-point Green functions are known to depend on the ensemble for
which they are calculated in a complicated way \cite{bpiz}.
They are completely non-universal.
 From a diagrammatic standpoint, the key difference
between Wigner and invariant ensembles is seen in the gluon interaction
vertices.
The diagram expansion entails summing over all possible ways of taking the
the averages of groups of matrices in an expression.
In the case of an invariant class problem, enumerating the diagrams to be
summed involves perturbatively expanding
the ensemble density around the Gaussian case \cite{bz3}.  The perturbations,
which
are identified with gluon interaction vertices in diagrams,
are invariant under the same transformation as the averaging ensemble.
Because of this, a $k$th order gluon interaction
vertex will have $k$ indices which are summed on rather than just the two
that an analogous Wigner type vertex would have.
 Such vertices may not be neglected in the large $N$ limit, and by their
inclusion
the one-point Green function becomes dependent on the distribution
in detail.
Diagrammatic representations of a simple four-point vertex are displayed
in Figure \ref{vertices} in order to contrast their index structures.

Returning to our Wigner class calculation, the diagram expansion
for $G(z)^i_j$ looks like
that in Fig \ref{series}(a), in the large $N$ limit.
By introducing $\Sigma(z)^i_j$ as the sum of irreducible diagrams---that
is, diagrams which may not be cut on a single quark line to produce two
complete and independent ones---we obtain the ``integral" equation,
\begin{equation}
\Sigma(z)^i_j = \sigma^2\left({1\over z- \Sigma(z)}\right)^i_j
\label{sigeqn}
\end{equation}
which is depicted in Fig. \ref{series}(b).  It is obvious from the graphs that
$\Sigma(z)^i_j = \delta^i_j \Sigma(z)$, thus  Eq. (\ref{sigeqn}) can be written
as
a simple quadratic equation. Solving this for
 $\Sigma$, and using the fact that $\Sigma(z) = \sigma^2 G(z)$
yields the Green function
\begin{equation}
G(z) = {1\over 2\sigma^2}(z - \sqrt{z^2-4\sigma^2})
\label{gfunc}
\end{equation}
This of course corresponds to the eigenvalue density Eq. (\ref{ssl}), a long
known result.  Since it only depends on
$\sigma^2$ and not on any other details of the ensemble distribution, it
holds for any ensemble in the Wigner class, and therefore is known as a
universal
result. In our further results, however, we will find an increasing dependence
 on the particulars of the ensemble distribution starting with
the two-point connected Green function.

\subsection{Two-point connected Green function}
We write $G_c(w,z)$ as an expansion in $1/z$ and $1/w$ with two derivatives
taken out explicitly.
\begin{eqnarray}
N^2 G_c(w,z)
& = & {1\over wz}  \sum_{n=1}^\infty\sum_{m=1}^\infty
        \left\langle\text{Tr}\left(\varphi{1\over w}\right)^n
        \text{Tr}\left(\varphi{1\over z}\right)^m\right\rangle_c \nonumber \\
& =
&{\partial\over\partial w}{\partial\over\partial z}
        \sum_{n=1}^\infty \sum_{m=1}^\infty {1\over n m}
        \left\langle\text{Tr}\left(\varphi{1\over w}\right)^n
        \text{Tr}\left(\varphi{1\over z}\right)^m\right\rangle_c,
\end{eqnarray}
where the subscript $c$ denotes the connected part.
In diagram language this expression is a sum of graphs that consist of two
different
quark loops interacting via one or more gluons. Notice that since we have taken
out
the two derivatives there are an equal number of $\varphi$'s and propagators
 in each trace.
The action of each derivative is to sum over all the possible insertions
of a quark-quark vertex on it's respective quark loop.
The evaluation of the diagrams
 will be made easier by having taken out these derivatives.

Consider first the case where there are no gluon-gluon interactions,
only quark-gluon ones (namely, the Gaussian case), and where contractions are
made
only between the two traces.  In this case then $m$ and $n$ must be equal.
If we draw the diagrams so that one loop is inside of another then
the planarity of graphs is easy to determine.
There are $n$ ways of connecting the first gluon from one
loop to another and then only one way of drawing the rest without crossing.
Each connection corresponds to averaging together a $\varphi$ in one trace with
a $\varphi$ in the other and results in a factor of $\sigma^2/N$.  In the
double line
formalism we see each gluon connection creates a new closed index loop
 and thus contributes a factor of $N$. These graphs sum to
\[
-{\partial\over\partial w}{\partial\over\partial z}\log\left(1-{\sigma^2\over
wz}\right).
\]

Next consider higher order gluon interactions, still only treating inter-trace
connections.  Each gluon-gluon interaction vertex of order $k$, corresponding
to the
averaging together of $k$ $\varphi$'s split between the traces,
 contributes a multiplicative
factor of $C_k/N^{k/2}$ while adding at most one additional summed index loop.
Since $G_c(w,z)$ is of order $N^{-2}$, in this case only one such graph is
allowed,
the one with four gluons, two on each loop, interacting via a four point vertex
$(k=4)$. This graph, depicted in Fig. \ref{4graph}(a), averages to
\[
{\tau^4 \over 2}{\partial\over\partial w}{\partial\over\partial z}
\left({1\over wz}\right)^2,
\]
with $\tau^4 \equiv C_4$, the fourth moment of the
ensemble distribution.

We must next consider contractions within the same trace.
These consist of what could be termed ``vertex" and ``self energy''
corrections.
By a vertex correction we refer to the inclusion of gluons
which connect two quark lines in the same loop,
on either side of a group of one or more quark-gluon vertices.
We immediately see that vertex corrections may be ignored since
such a gluon turns two closed index loops into one, suppressing the diagram
in which it appears by at least $N^{-2}$.
By self energy corrections
we refer to the inclusion of gluons emitted and re-absorbed between vertices.
The inclusion of self energy corrections to the quarks is instantly
achieved through the ``dressing" of the bare quark propagators $1/w$ and $1/z$
by taking $1/w \rightarrow G(w)$ and $1/z \rightarrow G(z)$ everywhere,
excepting the partial derivatives.

Thus, we obtain the complete two-point connected Green function for
Wigner class ensembles in general,
\begin{equation}
N^2 G_c(w,z) = -{\partial\over\partial w}{\partial\over\partial z}
        \left(\log\left[1-\sigma^2G(w)G(z)\right] - {\tau^4\over
2}G(w)^2G(z)^2\right).
\label{dsoln}
\end{equation}
For the case of Gaussian distributed matrix elements $\tau^4 = 0$ and Eq.
(\ref{dsoln})
reduces to the well known result for the Gaussian unitary
ensemble (GUE) as we
would expect.  In general though, we find a dependence
on two parameters, the second and fourth moments of the distribution.

\subsection{Three-point and higher Green functions}
\label{sec:3pthigher}

Here we will discuss the application of the diagram method to the calculation
of $n$-point connected Green functions.
In the previous section we saw that, before the
dressing procedure, the expansion for the general two-point function
differed from that for the GUE by only one diagram.
This diagram, once dressed, may be viewed as a correction to the GUE result.
As we will see, the higher
order connected functions lend themselves to a similar interpretation, and
may be written
\begin{equation}
G_{cN}(z_1,z_2,\ldots,z_n) = G_{cN}^{(0)}(z_1,z_2,\ldots,z_n)
        + (\text{non-Gaussian terms}),
\label{gcnform}
\end{equation}
where the superscript $(0)$ indicates the GUE result.
 From here on we will not explicitly calculate the GUE Green functions
as they are well known in the literature (see, for example, \cite{verb} or
\cite{eynard}).
  Our main interest will be on
the non-Gaussian terms and the diagrams that comprise them.

The relevant, undressed non-Gaussian diagrams are straightforward to enumerate.
By our conventions an $n$-point connected Green function for the GUE, as well
as
the diagrams in its associated expansion, is of order $N^{-2n+2}$.  One creates
the
non-Gaussian diagrams for a general ensemble by considering gluon
interaction vertices; these have a few qualities which greatly reduce the
number of relevant diagrams in the large $N$ limit.

First, a $k$th order gluon interaction vertex ($k>2$) contributes a
multiplicative
factor of $C_k/N^{k/2}$ while only having at most two free indices associated
with it which may
be summed on.  Including such a vertex in a connected diagram
makes the diagram order $N^{-n-k/2+2}$ or less.  This fact limits the moments
that
the $n$-point function can depend on to those up to $C_{2n}$.  As we have seen
the
2-point function depends only on $C_2$ and $C_4$, the
3-point function will depend on $C_2$,$C_4$ and $C_6$, {\it etc.}
This fact also limits the number
of vertices which may appear in any one diagram.  In
the 2-point case only a single $k=4$ vertex is allowed; the inclusion of
another
creates a diagram of order $N^{-4}$ or less.
In the three-point case we may have a single $k=6$ vertex, or up to two $k=4$
ones.

Second, any two quark loops connected via a gluon interaction vertex may not
have any
other connections between them to leading order.  The simplest case where this
may be observed
is in the single non-Gaussian diagram of the two-point function,
depicted in Fig.\,\ref{4graph}(a).  If we add a single gluon connecting the two
loops we obtain Fig.\,\ref{4graph}(c).  This new graph is now down by $N^{-1}$
because
we have added a gluon contributing $\sigma^2/N$ yet created no new sum loops
since
the gluon interaction vertex requires the lines on either side of it to have
the
same index value.  This effect persists in the general $n$-point connected
Green function.

As a demonstration of the use of these simplifying facts we calculate
 the three-point connected Green function to
leading order.  From our previous discussion we determine that the non-Gaussian
diagrams of
the three-point function will depend only on the $C_4$ and $C_6$ moments
and the diagrams in their expansion will have either one 6-point, one
four-point,
or two four-point gluon interaction vertices [see
Fig.\,\ref{3ptcorrections}(a)].

There is only one, undressed diagram containing a 6-point gluon interaction
vertex which is relevant to order $N^{-4}$.
It corresponds to a term
\begin{equation}
-{\upsilon^6\over 2 N^4}
\partial_1\partial_2\partial_3\left( {1\over z_1 z_2 z_3} \right)^2,
\label{6ptterm}
\end{equation}
with $\upsilon^6 \equiv C_6$.

In the case where we have a single four-point gluon interaction vertex we have
two
sub-cases which are not suppressed:
 one with the vertex connecting all three loops, and the other with the
vertex connection two of the three.  With the vertex connecting all three there
is only one basic diagram (plus permutations),
two gluons connecting to one quark loop and one each to the
other two.  This corresponds to,
\begin{equation}
-{\tau^4\over 2 N^4}\partial_1\partial_2\partial_3
        \left({1\over z_1 z_2 z_3}
        \left({1\over z_1}+{1\over z_2}+{1\over z_3}\right)\right)
\label{4ptterm1}
\end{equation}

If the four-point vertex only connects two of the quark loops there are an
infinite
number of graphs.
This occurs because, although once we have connected two of the three quark
loops via a four-point
gluon vertex there may be no other connections between them, either one or the
other of the loops may have any number of non-interacting (Gaussian) gluon
connections
to the third.  The resulting diagram expansion will be identical to the
expansion of the two-point connected Green function summed over all possible
ways to connect a third loop via a four-point gluon vertex.   We use the
distributive
property of the derivative to immediately write this as
\begin{equation}
- {\tau^4\over N^4} \partial_1\partial_2\partial_3
     \left\{ {1\over z^2_1}
        \left( {1\over z_2}\partial_2 + {1\over z_3}\partial_3 \right)
\left[ \log\left(1-{\sigma^2\over z_2 z_3}\right)
        - {\tau^4\over 2}{1\over z_2^2 z_3^2} \right]
        + (1 \leftrightarrow 2) + (1 \leftrightarrow 3)
        \right\}.
\label{4ptterm2}
\end{equation}
A note, relating to the interpretation of this problem as a field theory
calculation,
is that we would not be able to obtain this sum of diagrams so easily
if it were not for the fact we are dealing with a static, zero
dimensional field theory.
In a field theory with ``time" and ``space" variables, quark lines would carry
momentum which would be conserved at vertices.  Then the derivative trick
would no longer work.

Finally, one can see that the inclusion of diagrams with
two four-point gluon interactions is achieved by
substituting $G_{c23}^{\text{und.}}$ for
$G_{c23}^{(0)\text{und.}}$
in Eq.\,(\ref{4ptterm2}).  Putting all the terms together,
dressing them via $1/z_i\rightarrow G_i$, and simplifying some,
we find the three-point connected Green function is
\begin{eqnarray}
& N^4G_{c123} =  N^4G_{c123}^{(0)}
        -  \partial_1\partial_2\partial_3
\Biggl\{ \tau^4 G_1 G_2 G_3 (G_1+G_2+G_3) + {\upsilon^6\over 2} G^2_1G^2_2G^2_3
 \nonumber \\
&  + \tau^4 \left[ G_1^2 \left( G_2\partial_2+G_3\partial_3\right)
        \left(\log(1-\sigma^2G_2G_3) - {\tau^4\over 2}G_2^2G_3^2\right)
        + (1 \leftrightarrow 2) + (1 \leftrightarrow 3) \right]
\Biggr\}.
\label{gc3}
\end{eqnarray}

\subsection{$n$-point connected Green functions for $C_n \neq 0$}
\label{sec:cneqzero}

Following the procedure above one
may generate the entire hierarchy of connected Green functions.
In general,the $n$th connected Green function
will equal the GUE $n$-point function plus additional terms depending
the ensemble distribution's moments up to $C_{2n}$ and on powers of the
one-point
function and its derivatives. However, if the ensemble has an $n$th moment the
form of the $n$-point Green function will be far simpler.

Consider our previous example. In the three-point function
calculation above we specified an even distribution for the ensemble.
If we allow both even and odd moments of the ensemble distribution to exist we
must also consider for this case diagrams containing three-point gluon
interaction vertices.
We define a third moment of the distribution:
$\langle\varphi^i_j\varphi^j_i\varphi^i_j\rangle = ({1+i\over \sqrt{2}} +
(1-{1+i\over \sqrt{2}})\delta^i_j)C_3/N^{3/2}$
 (no sum on repeated indices).
Then, three additional, undressed diagrams, depicted in
Fig.\,\ref{3ptcorrections}(b),
contribute to the three-point connected Green function. The
second two are order $N^{-4}$---the same as all the original terms in
Eq.(\ref{gc3}).
The first, however, is order $N^{-7/2}$.  In other words, this single
diagram is order $\sqrt{N}$ larger than all the others.  Thus, we have
\begin{equation}
N^{7/2}G_{c123} =
-\varsigma^3\partial_1\partial_2\partial_3\left(G_1G_2G_3\right)
                                + {\sf O}(N^{-1/2})
\end{equation}
where $\varsigma^3 \equiv C_3$.

For an ensemble with a non-zero $n$th moment ($n > 2$) one finds that
the $n$-point connected Green function, to leading order, is
\begin{equation}
N^{3n/2-1}G_{c1\ldots n} = C_n\prod^n_{i=1}(-\partial_i)G_i
        + {\sf O}(N^{-1+1/2 (n\bmod 2)}).
\end{equation}
Thus, depending on the particulars of the ensemble density, the leading order
$n$-point connected correlation function can be a quantity of order
$N^{-3n/2+1}$
through order $N^{-2n - 2}$.
This sort of dependence
on the existence of various ensemble moments is in
striking contrast to the case in Invariant class ensembles.  For the
Invariant class, 2-point and higher
connected correlation functions depend on the ensemble density only through the
endpoint of the eigenvalue spectrum and the order in $1/N$ of the various
quantities is, in general, independent of the ensemble density.

\section{Renormalization group approach}
\label{sec:rg}
In this section we demonstrate the use of a renormalization group inspired
approach
\cite{bjj} to calculate the connected two-point Green function
for theories in the Wigner class. Further application to higher point
 functions follows in an obvious way.

This method has been used previously, by Br\'ezin and Zee \cite{bz2}, to derive
the
one-point green function for the Wigner class of theories, as well as by others
to
derive various quantities in invariant class theories \cite{sakai}.
In the interest of making the somewhat complicated derivation of the two-point
function more transparent we will briefly
review Br\'ezin and Zee's calculation of the one-point function.

\subsection{One-point Green function via RG}

The basic idea here is to write
the ($N+1$) dimensional, matrix Green function in terms of dimension
$N$ quantities and average over select elements. This generates a partial
differential
equation which may then be solved in the large $N$ limit.

To begin the derivation of $G(z)$ we write the $(N+1)\times(N+1)$ Hermitian
matrix $\varphi_{N+1}$ in the form
\[
\varphi_{N+1} = \left( \begin{array}{cc}
                                \varphi_N & v \\
                                v^{\dag}   & \chi
                        \end{array} \right),
\]
with $\varphi_N$ an $N\times N$ Hermitian matrix,  $v$ a complex $N$ component
vector,
and $\chi$ a real scalar then the identity
\begin{equation}
\text{Tr} {1\over z-\varphi_{N+1}} = \text{Tr} {1\over z-\varphi_N}
        + {1+\langle v|\left({1\over z-\varphi_N}\right)^2|v\rangle \over
        z-\chi-\langle v|{1\over z-\varphi_N}|v\rangle}  \label{ident}
\end{equation}
may be established.  One can write this as a partial differential equation for
$\hat{G}$,
\begin{equation}
\text{Tr} \hat{G}_{N+1}(z) = \text{Tr} \hat{G}_N(z)
        + {\partial \over \partial z}
        \log(z-\chi-\langle v|\hat{G}_N(z)|v\rangle). \label{ghatpde}
\end{equation}
We obtain a differential equation for $G(z)$ by averaging Eq. (\ref{ghatpde})
over the
distribution governing $\varphi_{N+1}$, and
retaining terms of order $N^0$ or larger.

The term on the left hand side of Eq. (\ref{ghatpde}) averages to
\[
(N+1)G_{N+1}(z) = (N+1)G_N(z) + (N+1){\partial \over \partial N}G_N(z) + \ldots
\]
Since $G_N(z)$ is of order $N^0$, to the desired order we may neglect all but
the first
term on the right.

Next, we skip to the second term on the right of Eq. (\ref{ghatpde}). Recall
that the mean squared value of matrix elements is of order $N^{-1}$. To
order $N^0$, then, we may immediately set $\chi = 0$ in this term.
If we expand the logarithm we
seen that we will be required to evaluate averages such as
$\langle v^*_iv_jv^*_kv_l\ldots v^*_pv_q\rangle$.  Consider the term with
four $v$'s
\[
\langle v^*_iv_jv^*_kv_l\rangle =
A(\delta_{ij}\delta_{kl}+\delta_{il}\delta_{kj})+B\delta_{ijkl}
\]
In the large $N$ limit $A$ terms will dominate, with
$A = \langle v^*_1v_1v^*_2v_2\rangle
   = \langle v^*_1v_1\rangle^2 = (\sigma^2/N)^2$
In general every average of a product of $v$'s will be dominated by
second moment terms, and in fact we may replace
$\langle v| 1/(z-\varphi_N)|v\rangle$ with $(\sigma^2/N) G_N(z)$
within the logarithm.

We must be more careful in averaging the first term on
the right of Eq. (\ref{ghatpde})
since we are averaging over the distribution appropriate for $\varphi_{N+1}$,
and not $\varphi_N$.  Our experience with the second term, which we treated
in the previous paragraph, indicates how free our choice of a matrix density
may be.  First, $G(z)$ depends only on the second moment, $\sigma^2$, of
the distribution.  In fact we may set off diagonal elements to zero
with some finite probability without changing $G(z)$.
Second, diagonal and off
diagonal elements may even obey different distributions since all that is
required by this procedure is that
$\langle \chi^2 \rangle \rightarrow 0$ as $N \rightarrow \infty$.

Thus, if $P(\varphi_{ij};\sigma^2/N,\varsigma^3/N^{3/2},\tau^4/N^2,\ldots)$,
defines a particular distribution (with $\sigma^2$,
$\varsigma^3$,
$\tau^4$, $\ldots$ being its moments)
then it is sufficient to use $P(\varphi_{ij};\sigma^2/N,0,0,\ldots)$ in the
calculation
of $G(z)$, that is, we may effectively set $\varsigma$,$\tau$,$\ldots$
 equal to zero.
After some straightforward manipulation we see that first term on
the right of Eq. (\ref{ghatpde}) averages to $N G\left[ z;\left(1-{1\over
N}\right)\sigma^2 \right] + {\sf O}(N^{-1})$.

Putting all this together, we find that in the limit $N\rightarrow\infty$,
Eq. (\ref{ghatpde}) averages to
\begin{equation}
G(z) = - \sigma^2{\partial\over\partial\sigma^2} G(z)
        + {\partial\over\partial z}\log(z-\sigma^2G(z))
\end{equation}
 From dimensional analysis one can see that
$G(z;\sigma^2) = (1/\sigma^2) G(z/\sigma;1)$ and hence
$\sigma^2{\partial\over\partial\sigma^2} G(z)
= -\frac{1}{2} (1+z{\partial\over\partial z}) G(z)$.  Finally, we find
that Eq. (\ref{ghatpde}) becomes
\begin{equation}
G(z) = z {d\over dz} G(z) +  2 {d\over dz}\log(z-\sigma^2 G(z)).
\end{equation}
This nonlinear equation has the unique solution
\begin{equation}
G(z) = {1 \over 2\sigma^2} (z - \sqrt{z^2 - 4\sigma^2}). \label{g}
\end{equation}
Naturally, this $G(z)$ is identical to that in Eq. (\ref{gfunc}) and thus
corresponds to the eigenvalue density (\ref{ssl}).
Now we will apply this method to the problem of deriving the
connected two-point Green function.

\subsection{Two-point connected Green function via RG}
\label{sec:2pt}

As a starting point, we use Eqs. (\ref{ghat}) and (\ref{ident}) to write
\begin{eqnarray}
\text{Tr}\,\hat{G}_{N+1}(w)
        \text{Tr}\,\hat{G}_{N+1}(z)
        & = &\text{Tr}\,\hat{G}_N(z)\text{Tr}\,\hat{G}_N(w) \nonumber \\
        &   & +{\partial\over\partial w}
        \text{Tr}\,\hat{G}_N(z)\log(w-\chi-\langle v|\hat{G}_N(w)|v\rangle)
\nonumber       \\
        &   & +{\partial\over\partial z}
        \text{Tr}\,\hat{G}_N(w)\log(z-\chi-\langle v|\hat{G}_N(z)|v\rangle)
\nonumber       \\
        &   & +{\partial^2\over\partial w\partial z}
        \log(w-\chi-\langle v|\hat{G}_N(w)|v\rangle)
        \log(z-\chi-\langle v|\hat{G}_N(z)|v\rangle).
\label{treqn}
\end{eqnarray}
Now we must average over the distribution governing $\varphi_{N+1}$, keeping
only {\it
connected} terms up to order $N^{-1}$.

The term on the left of Eq. (\ref{treqn}) is trivially averaged to
\[
(N+1)^2 G_{c\,(N+1)}(w,z) = (N+1)^2 G_{c\,N}(w,z) +
(N+1)^2{\partial\over\partial N}G_{c\,N}(w,z) + \ldots
\]
Since we are keeping only terms up to order $N^{-1}$, and we know that
$G_{c\,N}(w,z)$
itself is order $N^{-2}$, then we may write this as
\begin{equation}
N^2 G_{c\,N}(w,z) + {\sf O}(N^{-2})
\label{lhs}
\end{equation}

Next we consider the second (and third) term
on the r.h.s. of Eq. (\ref{treqn}).
Let us first obtain the connected average to leading order in $1/N$.
To leading order we may immediately
set $\chi=0$ and take the average over $v$ inside the logarithm
where we see we may again replace
$\langle v|\hat{G}_N(w)|v\rangle$ with
$(\sigma^2/N)\text{Tr}\,\hat{G}(w)$.
Thus, we have
\begin{equation}
-{\partial\over\partial w}\left\langle\text{Tr}\hat{G}_N(z)
        \sum_{n=1}^{\infty}\frac{1}{n}\left(\frac{1}{w}\frac{\sigma^2}{N}
        \text{Tr}\hat{G}_N(w)\right)^n \right\rangle_c
\end{equation}
Now we complete the average over the remaining variables, here using the
distribution appropriate to $\varphi_N$ since corrections are of order
$N^{-1}$ and still only concerning ourselves with leading order contributions.
To leading order we need only average together $\text{Tr}\hat{G}(z)$
and one of the $\text{Tr}\hat{G}(w)$'s in each series term.
Then $G_N(w)$ may be substituted for the remaining $\text{Tr}\hat{G}(w)$'s.
By summing the series, and utilizing the identity
\[
G_N(w) = {1 \over w-\sigma^2G_N(w)} + {\sf O}(N^{-1}),
\]
we obtain
\begin{equation}
-N{\partial\over\partial w}\sigma^2G_{c\,N}(w,z)G_N(w)
\label{t2}
\end{equation}
as the leading order average of the term in question.  Note that this
expression is already order $N^{ - 1 }$, therefore we need not consider any
corrections to it as they will be order $N^{-2}$ or smaller.

On to the fourth term in Eq. (\ref{treqn}).  To the desired order
we retain at most a single second moment in $\chi$, any higher moments or
powers of
moments may be neglected, yielding
\begin{eqnarray}
{\partial^2\over\partial w\partial z} \left[
 \left\langle\sum_{m,n=1}^\infty\frac{1}{mn}
        \left({1\over w}\langle v|\hat{G}_N(w)|v\rangle\right)^n
        \left({1\over z}\langle v|\hat{G}_N(z)|v\rangle\right)^m
\right\rangle_c \right. \nonumber \\
 \left. + \frac{1}{N}{\sigma^2\over zw}
  \left\langle
      \left( 1 + \sum_{n=1}^\infty\frac{1}{n}{1\over w^n}
                \langle v|\hat{G}_N(w)|v\rangle^n
      \right)
      \left( 1 + \sum_{m=1}^\infty\frac{1}{m}{1\over z^m}
                \langle v|\hat{G}_N(w)|v\rangle^m
      \right)
  \right\rangle
\right]         \label{t4a}
\end{eqnarray}

In the first term of (\ref{t4a}) contributions to order $N^{-1}$ will come
from one of two connected averages: either
\begin{eqnarray}
\langle v_iv_l^*\rangle\langle v_kv_j^*\rangle\hat{G}(w)^i_j\hat{G}(z)^k_l
& = &\frac{\sigma^4}{N^2} \text{Tr}\left[\hat{G}(w)\hat{G}(z)\right],
\text{ or} \label{avga} \\
\langle v_iv_l^*v_kv_j^*\rangle_c\hat{G}(w)^i_j\hat{G}(z)^k_l
& = &\frac{\tau^4}{N^2} \sum_{i=1}^N\hat{G}(w)^i_i\hat{G}(z)^i_i
\label{avgb}
\end{eqnarray}
where $\tau^4$ is the fourth moment of the ensemble distribution.
Each term is of order $N^{-1}$, thus either one or the other will occur once.
The contribution from (\ref{avga}) is
\begin{eqnarray}
{1\over N} {\partial^2\over\partial w\partial z}
   \left[
        {1\over w- \sigma^2G(w)}{1\over z- \sigma^2G(z)}
        \left\langle
                {\sigma^4\over N}\text{Tr}(\hat{G}_N(w)\hat{G}_N(z))
        \right\rangle
   \right]      \nonumber \\
 =  - {1\over N} {\partial^2\over\partial w\partial z}
        \left[ {G(w)-G(z)\over w-z} \right]  \sigma^4 G(z) G(w) \nonumber \\
 =  {1\over N}{\partial^2\over\partial w\partial z}
        { \sigma^4G^2(w)G^2(z)\over 1- \sigma^2G(w)G(z)},
\label{t4b}
\end{eqnarray}
and the contribution from (\ref{avgb}) is
\begin{equation}
{1\over N} {\partial^2\over\partial w\partial z}\left[
{1\over w- \sigma^2G(w)}{1\over z- \sigma^2G(z)}
  \left\langle
     {\tau^4\over N}\sum_{i=1}^N\hat{G}_N(w)^i_i\hat{G}_N(z)^i_i
  \right\rangle \right]
  = {\tau^4\over N}G^2(w)G^2(z)
\label{t41}
\end{equation}

Treating the second term in (\ref{t4a}) is far simpler since its average is
unconnected
and it is manifestly of order $N^{-1}$.
We may immediately average it to
\begin{equation}
{1\over N} {\sigma^2\over zw}
        \left[1+{{\sigma^2\over w}G(w)\over 1-{\sigma^2\over w}G(w)}\right]
        \left[1+{{\sigma^2\over z}G(z)\over 1-{\sigma^2\over z}G(z)}\right]
= {\sigma^2 \over N} G(w) G(z)
\label{t42}
\end{equation}

Finally, we come to the first term on the r.h.s. of Eq. (\ref{treqn}),
 we have a case where, as in the calculation of $G(z)$,
 we must carefully average over the distribution appropriate to
$\varphi_{N+1}$ and not $\varphi_N$.
Again we are led to consider how free our choice of probability distribution
may be.  Consideration of the previous terms has shown that $G_{c\,N}(w,z)$
will
explicitly depend on only two numbers $\sigma^2$ and $\tau^4$, thus it is
sufficient to average this term using the general distribution
$P(\varphi_{ij};\sigma^2/N,\tau^4/N^2)$.

We find the term in question then averages to
\[
\left[ 1 - {1\over N}\left(\sigma^2{\partial\over\partial\sigma^2}
        + 2 \tau^4{\partial\over\partial\tau^4}\right)\right]G_{c\,N}(w,z)
\]
 From dimensional analysis one can see that
\[
G_{c\,N}(w,z;\sigma^2,\tau^4)={1\over \sigma^2}
G_{c\,N}(w/\sigma,z/\sigma;1,\tau^4/(\sigma^2)^2).
\]
Hence, the identity
\[
\sigma^2{\partial\over\partial\sigma^2}
        G_{c\,N}(w,z) = -\left(1+{z\over 2}{\partial\over\partial z}
                +{w\over 2}{\partial\over\partial w}
                + 2\tau^4 {\partial\over\partial\tau^4}\right)
        G_{c\,N}(w,z).
\]
Using this we see that the first term on the r.h.s. of Eq. (\ref{treqn})
finally
averages to
\begin{equation}
\left(1 + {1\over 2}{\partial\over \partial w} w
        + {1\over 2}{\partial\over  \partial z} z \right)
         N^2 G_{c\,N}(w,z) + {\sf O}(N^{-2})  \label{t12}
\end{equation}

Using expressions (\ref{lhs}), (\ref{t2}), (\ref{t41}), (\ref{t42})
and (\ref{t12}) we obtain the partial differential equation
\begin{equation}
\left[\sum_{i=1,2}\partial_i(z_i-2\sigma^2G_i)\right]
        N^2G_{c12} = -2\partial_1\partial_2
        \left({\sigma^2G_1G_2\over1-\sigma^2G_1G_2}+\tau^4G_1^2G_2^2\right)
+ {\sf O}(1/N),
\label{gpde}
\end{equation}
where we have used the
obvious shorthand: $G_{c12}\equiv G_c(w,z)$,
$\partial_1\equiv (\partial/\partial w)$,
$G_1\equiv G(w)$, {\it etc}.
Note that the derivatives on the l.h.s. act on $G_{c12}$ as well as on
the parenthesized terms immediately to their right.

This rather complicated, non-linear equation is solvable.
In general we may write its solution in the form
\begin{equation}
N^2 G_{c12} = \partial_1\partial_2 F(G_1+G_2,G_1-G_2) + h_1(G_1) + h_2(G_2)¸
\end{equation}
with $F$, $h_1$, and $h_2$ unknown functions.
We may immediately set $h_{1,2} =0$ since
$G_{c12}$ must go like $1/z_{1,2}^2$ as $z_{1,2}\rightarrow\infty$.
We substitute this general solution into Eq. (\ref{gpde}) and use
the large $N$ identity,$G/G^\prime= 2\sigma^2G - z$,
where $G^\prime\equiv(\partial/\partial z) G(z)$,
\begin{equation}
\partial_1\partial_2\left[\left(
        {G_1\over G_1^\prime}\partial_1+{G_2\over G_2^\prime}\partial_2
        \right) F(G_1+G_2,G_1-G_2)\right]
        = \partial_1\partial_2
        \left[{2\sigma^2G_1G_2\over1-2\sigma^2G_1G_2}
         +\tau^4G_1^2G_2^2\right].
\label{subspde}
\end{equation}
The structure of the equation suggest that we make the variable substitution
$g_\pm \equiv\frac{1}{\sqrt{2}}(\log G_1\pm\log  G_2$). We equate the bracketed
terms in
Eq. (\ref{subspde}) to obtain
\begin{equation}
\sqrt{2}{\partial\over\partial g_+} F(e^{g_+},e^{g_-})
        = {2\sigma^2e^{g_+}\over 1-\sigma^2e^{g_+}} + 2\tau^4e^{2g_+},
\end{equation}
which is trivially integrable.  Thus,
\begin{equation}
N^2 G_{c12} = -\partial_1\partial_2\left[\log(1-\sigma^2G_1G_2)
         - {\tau^4\over 2}G_1^2G_2^2 \right] + f(G_1/G_2)
\end{equation}
where $f$ is an unknown function.
The known asymptotic behavior of $G_{c12}$  requires $f=0$, thus the
unique solution to Eq. (\ref{gpde}) is
\begin{equation}
N^2 G_c(w,z) = -{\partial\over\partial w}{\partial\over\partial z}
        \left(\log\left[1-\sigma^2G(w)G(z)\right] -{\tau^4\over
2}G(w)^2G(z)^2\right)
\label{gsoln}
\end{equation}
By Eq.(\ref{2ptdif}) we find the corresponding correlation function
is
\begin{equation}
 \rho_c(\mu,\nu) = {-1\over 4\pi^2N^2}{\partial\over\partial
\mu}{\partial\over\partial \nu}
        \left( \log{4\sigma^2 - \mu\nu +
\sqrt{(4\sigma^2-\mu^2)(4\sigma^2-\nu^2)}
                \over 4\sigma^2 - \mu\nu -
\sqrt{(4\sigma^2-\mu^2)(4\sigma^2-\nu^2)}}
    -{\tau^4 \mu \nu\over 2\sigma^8}
        \sqrt{4\sigma^2-\mu^2}\sqrt{4\sigma^2-\nu^2}\right),
\label{2ptd}
\end{equation}
which corresponds, when differentiated, to the previously displayed Eq.
(\ref{2ptfunc}).
We simplify by a change of variables to $\sin\theta \equiv \mu/2\sigma$,
$\sin\phi \equiv \nu/2\sigma$  which allows us to write
\begin{equation}
\rho_c(\theta,\phi) =  {-1\over 16\pi^2\sigma^2N^2 \cos\theta \cos\phi }
        {\partial\over\partial\theta}{\partial\over\partial\phi}
        \left( \log{1+\cos(\theta+\phi) \over 1 - \cos(\theta-\phi)}
                - {2\tau^4\over  \sigma^4}\sin 2\theta \sin 2 \phi \right)
\label{2ptdang}
\end{equation}
We can compare this to the known result for invariant class Hermitian ensembles
\begin{equation}
\rho_c(\theta,\phi) = {-1\over 4\pi^2a^2N^2\cos\theta \cos\phi }
        {\partial\over\partial\theta}{\partial\over\partial\phi}
         \log{1+ \cos (\theta+\phi) \over 1 - \cos (\theta-\phi) }
\label{2pttr}
\end{equation}
where $a$ is the endpoint of the eigenvalue spectrum and depends in detail on
the
ensemble distribution.  The Wigner class result is slightly ``less universal"
in
that it has an additional term and depends on a second number, $\tau^4$. As
we have seen, higher order connected Wigner correlation functions will depend
on
more and more moments of their ensemble distribution.

\section*{Conclusion}
For Wigner class ensembles, we have
 derived a general form
for the one, and two-point connected correlation functions using
both a diagrammatic and a renormalization group approach.
Using the diagrammatic approach, we have also shown how one may construct
higher order correlation functions,  in the large $N$ limit. By the RG inspired
method
we have demonstrated that results for the one and two-point functions are
unchanged
by considering some general cases where
the elements of matrices in the ensemble are not identically distributed. For
instance,
one may set elements to zero with some finite probability or allow on and off
diagonal elements
to have different distributions without changing the final result.

In general the
$n$-point function will depend on the ensemble's moments up to $C_{2n}$
and be comprised of
products and derivatives of the universal one-point Green function.
We have found, though,  in the case where a particular ensemble
distribution has a non-zero $n$th moment the
$n$-point connected Green function is particularly simple to leading order.
This contrasts
with the case for invariant class ensembles, where the one-point Green function
depends
on the ensemble in detail while higher order connected Green functions are
universal
in form.

\acknowledgements
We thank Roland Speicher for calling our attention to the work of Khorunzhy,
{\sl et.al.} in Ref. \cite{past}.
One of us (AZ) thanks E. Br\'ezin for discussions on this and other related
issues in random matrix theory. We are grateful to Joshua Feinberg for reading
this manuscript.
This work was supported in part by the National Science
Foundation under Grant No. PHY 89-04035.


%
%

\begin{figure}
\caption{The function $(\mu-\nu)^2\rho_c(\mu,\nu)|_{\mu=0}$ for the Wigner
class with parameters $\beta=1$, $\sigma^2 = 1$, and $N=100$, plotted
 in the two cases $\tau^4=0$ (solid line)  and $\tau^4 = -2$ (dashed line).}
\label{theory1}
\end{figure}

\begin{figure}
\caption{The function
$(\mu-\nu)^2\rho_c(\mu,\nu)|_{\mu=1}$ for the Wigner class with parameters
$\beta=1$, $\sigma^2 = 1$, and $N=100$, plotted in the two cases
 $\tau^4=0$ (solid line)  and $\tau^4 = -2$ (dashed line).}
\label{theory2}
\end{figure}

\begin{figure}
\caption{
Fig. (19) from Ref. \protect\cite{khkz}. Dashed line is a
 numerically obtained smoothed
connected correlation $(\mu-\nu)^2\rho_c(\mu,\nu)|_{\mu=0}$ for a real
symmetric Wigner class ensemble with matrix elements chosen to be
$\protect\pm 1/\protect\sqrt{N}$ with equal probability, and $N=100$.
Solid line is $(\mu-\nu)^2$ times the universal connected correlation
for real symmetric invariant class ensembles with $N=100$, and spectral
endpoints $\protect\pm 2$, at $\mu = 0$.}
\label{khkz1}
\end{figure}

\begin{figure}
\caption{
Fig. (20) from Ref. \protect\cite{khkz}. Dashed line is a numerically obtained
smoothed connected correlation $(\mu-\nu)^2\rho_c(\mu,\nu)|_{\mu=1}$ for a real
symmetric Wigner class ensemble  with matrix elements chosen to be
$\protect\pm 1/\protect\sqrt{N}$ with equal probability, and  $N=100$.
Solid line is $(\mu-\nu)^2$ times the universal connected correlation for real
symmetric invariant class ensemble, in the case $N=100$, and spectral endpoints
 $\protect\pm 2$, at $\mu=1$.}
\label{khkz2}
\end{figure}

 \begin{figure}
 \caption{The double line diagrams for terms in the expansion of $G(z)$
        which have four $\varphi$'s in them.}
 \label{4ds}
 \end{figure}

\begin{figure}
\caption{Wigner class versus invariant class vertices.}
\label{vertices}
\end{figure}

 \begin{figure}
 \caption{Diagram expansions in the limit of large $N$ for: (a) the one-point
        Green function, and (b) the sum of one-point irreducible graphs.}
 \label{series}
 \end{figure}

\begin{figure}
\caption{Various connected quark loops with interacting gluons.  Graph (a) is
        of order $N^0$, while (b) and (c) are of order $N^{-1}$, and (d) is
$N^{-2}$. }
\label{4graph}
\end{figure}

\begin{figure}
\caption{Graphical representation of the expansion of the ``undressed"
two-point connected Green function}
\label{gsum}
\end{figure}

\begin{figure}
\caption{ (a) Diagrams representing the undressed Gaussian
3-point connected Green function for an even ensemble distribution, and (b)
additional diagrams arising when odd moments are allowed.}
\label{3ptcorrections}
\end{figure}


%
%


\begin{references}
\bibitem{note} We choose this language for its convenience and familiarity
but the reader should keep in mind that there is no $U(N)$ symmetry
here so our use of the term ``gluon" is somewhat casual.
Also, what we, for simplicity's sake, call a $k$-point gluon interaction vertex
might more precisely be called an un-amputated, $k$-point gluon Green function.
\end{references}
\end{document}